\documentclass[10pt,floatfix,a4paper,sans,twocolumn,superscriptaddress,aps,pra,showkeys,notitlepage,showpacs,footinbib,eprint]{revtex4-1}
\usepackage{t1enc}
\usepackage[english]{babel}
\usepackage{amsmath,amssymb}
\usepackage{lipsum} 
\usepackage[utf8]{inputenc}
\usepackage{setspace}
\usepackage{array}
\usepackage[T1]{fontenc}
\usepackage{subfigure}
\usepackage{graphicx}
\graphicspath{{.}{figs/}}
\usepackage{dcolumn}
\usepackage{bbold}
\usepackage{bbm}
\usepackage{physics}
\usepackage{bm}
\usepackage{comment}
\usepackage[usenames]{xcolor}
\usepackage[unicode=true,pdfusetitle,
bookmarks=true,bookmarksnumbered=false,bookmarksopen=false,breaklinks=false,pdfborder={0 0 1},backref=false,colorlinks=true]
{hyperref}
\usepackage{accents}

\usepackage{times}
\usepackage{IEEEtrantools}
\begin{document}
	\title{Effective model for $A _{2g}$  Raman signal in URu$_2$Si$_2$}
	\author{Carlene Silva de Farias}
	\email{carlenepaula@gmail.com}
	\affiliation{Instituto de F\'{i}sica Gleb Wataghin, Unicamp, 13083-859, Campinas-SP, Brazil}
	\affiliation{Departamento de Física Teórica e Experimental and \\ International Institute of Physics, Universidade Federal do Rio Grande do Norte, Natal, Brasil.}
	\affiliation{Université Bordeaux, CNRS, LOMA,
		UMR 5798, F 33400 Talence, France}
	
	\author{Marie-Aude M\'easson}
	\affiliation{Institut N\'eel, CNRS/UGA UPR2940, 25 rue des Martyrs BP 166, 38042 Grenoble cedex 9, France}
	
	\author{Alvaro Ferraz}
	\affiliation{Departamento de Física Teórica e Experimental and \\ International Institute of Physics, Universidade Federal do Rio Grande do Norte, Natal, Brasil.}
	
	\author{Sébastien Burdin.}
	\email{sebastien.burdin@u-bordeaux.fr}
	\affiliation{Université Bordeaux, CNRS, LOMA,
		UMR 5798, F 33400 Talence, France}
	\date{\today}
	
\begin{abstract}
We propose an effective model to describe the $A_{2g}$ signal in Raman scattering experiments in the URu$_{2}$Si$_{2}$ compound. We follow the scheme proposed earlier by Khveshchenko and Wiegmann [\href{https://journals.aps.org/prl/abstract/10.1103/PhysRevLett.73.500}{Phys. Rev. Lett. \textbf{73}, 500 (1994)}] to calculate the $A_{2g}$ scattering vertex.
We extract the imaginary part of a two-point current-current correlation function and compare it directly with the Raman response. We obtain an inelastic peak at the $A_{2g}$ channel owing to the interplay between a local staggered ordering and a possible quantum spin liquid behavior. 
Our results offer an explanation for the electronic Raman scattering experiments at the hidden order phase in URu$_{2}$Si$_{2}$ compound [\href{https://journals.aps.org/prl/abstract/10.1103/PhysRevLett.113.266405}{Phys. Rev. Lett. \textbf{113}, 266405 (2014)}].
\end{abstract}


\maketitle

\section{Introduction \label{section1}}

The hidden order (HO) state of the heavy fermion compound URu$_2$Si$_2$ takes place at the critical temperature of $T_{0}\approx17.5$ K \cite{Palstra1985, Maple1986,Mydosh2011}.
Measurements of bulk thermodynamic and transport properties \cite{Broholm1987, Maple1986, Palstra1985} reveal that this state arises due to a second order phase transition from an initial paramagnetic state. According to the Landau classification for second order phase transitions, there must be an order parameter that assumes a nonzero value in the new state, as well as a symmetry that must be spontaneously broken that is associated with this transition. However, until today it has not been discovered which order parameter  or symmetry are associated with this state, and, in view of that, it has been called the hidden order phase. Many theoretical propositions as well as a vast arsenal of experimental techniques have been proposed and employed to investigate and characterize the HO states over the years. However, the question of what kind of spontaneous symmetry breaking and order parameter characterize this phase transition remain without answer \cite{Mydosh2011, Knafo2017}. 

Recently, Raman scattering has been one of the new experimental probes employed to investigate the HO phase \cite{Buhot2014, Kung2015}. Raman scattering proved to be a powerful experimental technique to understand and characterize the physics of strongly correlated systems \cite{Devereaux2007}. This technique has been used in the investigation and characterization of elementary excitations in high temperatures superconductors \cite{Shastry1990, Nagaosa1991, Khveshchenko1994, Devereaux1999}, in iron-based compounds \cite{Gallais2014, Gallais2016}, in heavy fermions systems \cite{Cooper1987, Buhot2014, Kung2015}, and in two-dimensional spin liquids \cite{Nasu2016, Glamazda2016}. 

Ref.\cite{Buhot2014,Kung2015} found that the HO is characterized by a sharp excitation at $1.7$ meV and a gap at the electronic continuum below $6.8$ meV. These signatures are of pure $A_{2g}$ symmetry. Buhot \emph{et al.} \cite{Buhot2014} concluded that the presence of this signal is compatible with a scenario in which the Brillouin zone (BZ) is folded from a body-centered tetragonal (BCT) to a simple tetragonal (ST) lattice taking place at the HO transition. Such scenario was also shown to be consistent with data from angle resolved photoemission spectroscopy \cite{Boariu2013, Bareille2014}.  Additionally, Kung \textit{et al.} \cite{Kung2015} also reported the presence of an additional signal in the $A_{1g}$ channel. They argue that the signal in the $A_{1g}$ channel is the results of a spontaneous symmetry breaking occurring at the HO phase transition on the crystal field scheme of the Uranium site. They then conclude that the order parameter has chiral nature.  They define this new ordered state as a commensurate hexadecapolar phase \cite{Harima2011}. These two studies provide new limitations to the current theories to explain the HO and prompt to the urgency of having a theoretical analysis based on Raman experiments.

In this paper, we propose an effective tight-binding model for URu$_{2}$Si$_{2}$ that explicitly considers the interplay between a spin liquid and a local staggered order phase. The existence of stable antiferromagnetic (AFM) and spin liquid (SL) phases in a BCT lattice structure with direct implications for URu$_{2}$Si$_{2}$ was previously discussed  by Farias \textit{et al.} \cite{Farias2016}. Both states display lattice translation symmetry breaking from BCT down to ST. Here, by using the symmetry classification proposed by Harima \textit{et al.} \cite{Harima2010}, we relate each phase with a specific spatial symmetry group. In order to compare our theory with experiments, we compute a two-point current-current correlation function in the $A_{2g}$ symmetry. We construct the $A_{2g}$ vertex by implementing the scheme put forward by Khveshchenko and Wiegmann \cite{Khveshchenko1994} in the context of  the high energy Raman scattering in Mott insulators system. Here, following those authors, we assume that the energies of the incident and scattered light are much larger than the widths of relevant electronic bands that, by itself, allow us to assume a resonant process. With this assumption, the Raman cross section is written in terms of the ground state correlation function of the scattering tensor $\hat{M}^{A_{2g}}$. This scattering tensor is proportional to the current-current response function. 

Our findings show that, as we tune the parameters that produce lattice symmetry breaking, we reproduce the sharp peak in the $A_{2g}$ symmetry
at $\omega\approx1.7$ meV, and this is only verified if we necessarily allow for the interplay between the local staggered order and a spin liquid state. The tail of the sharp peak starts to appear at $\omega\sim 2\Delta_{0}$, where $\Delta_{0}$ is the parameter that defines a local staggered order, and its intensity is controlled by the magnitude of the spin liquid parameter. This result is in reasonable agreement with the sharp response observed at $A_{2g}$ symmetry in the URu$_2$Si$_2$ \cite{Buhot2014}. Our work also connects to the earlier studies by Pepin \textit{et al.} \cite{Pepin2011} and Thomas \textit{et al.} \cite{Thomas2013} and offers a natural way of integrating the HO and AFM effects already in a realistic localized treatment.

The outline of the paper is as follows. Section \ref{secII} presents our model. We introduce the effective Hamiltonian with the specific parameters that take into account the possible phases associated with a specific space group. Section \ref{secIII} describes how we can obtain the signal for the $A_{2g}$ symmetry and presents also our numerical results. We finish this section by discussing the connection between the real experimental findings for URu$_2$Si$_2$ at $A_{2g}$ symmetry and our theoretical results and the spin liquid scenarios of our model. Finally, Section \ref{secIV}  presents our conclusion and summarizes our most significant results and their relation to other experiments and related systems.

\section{The Model \label{secII}}

\begin{figure}[!t]
	\subfigure[]{\includegraphics[height=4.0cm]{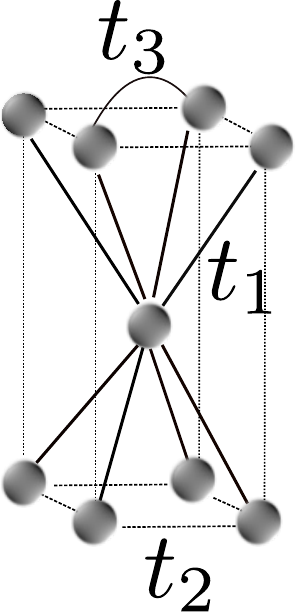}
		\label{hopping}}
	\quad 
	\subfigure[]{\includegraphics[height=3.5cm]{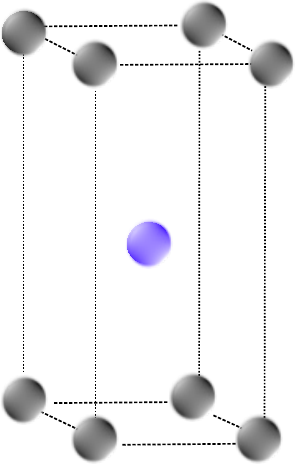}
		\label{AFM}}
	\quad
	\subfigure[]{\includegraphics[height=3.5cm]{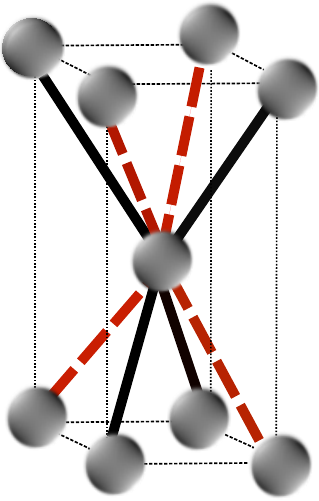}
		\label{msl}}
	\subfigure[]{\includegraphics[height=3.5cm]{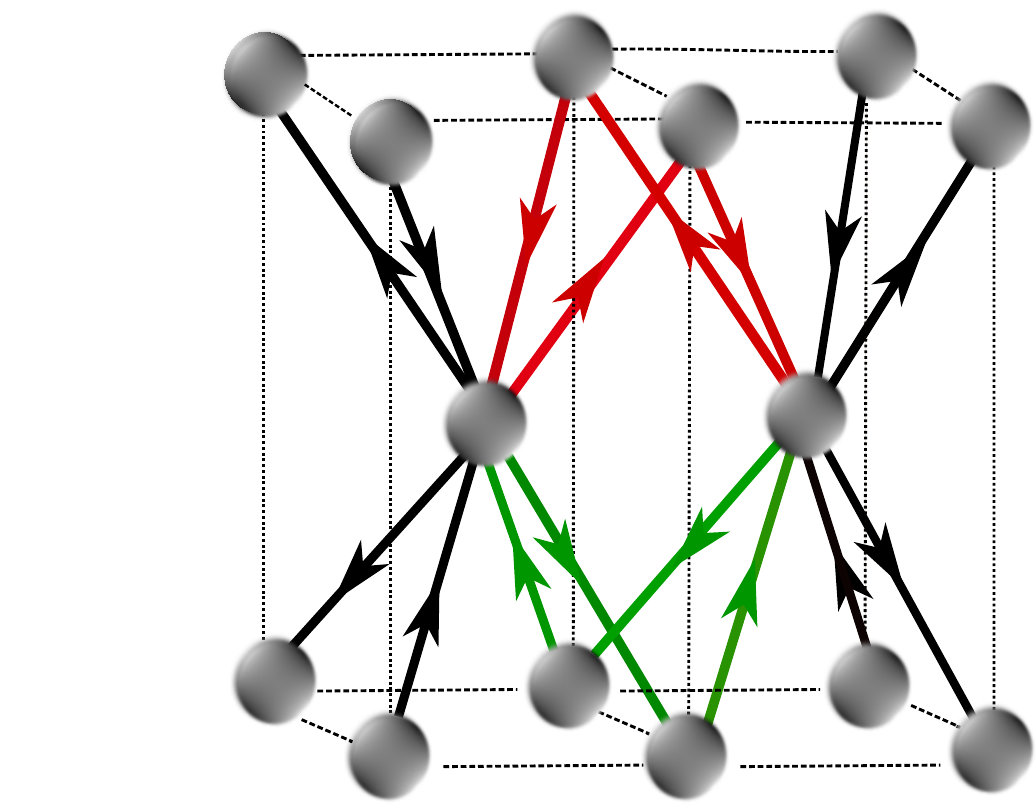}
		\label{csl}}
	\caption{\small{Schematic representation of the four possible ordered states that we consider in our model. (\textbf{a}) Paramagnetic state with full BCT lattice structure, space group No 139. (\textbf{b}) Antiferromagnetic state with simple tetragonal lattice, space group No 123. (\textbf{c}) Modulated spin liquid space group No 126. (\textbf{d}) Chiral spin liquid phase  space group No 128.}}
	\label{fig1}
\end{figure}
\begin{table*}
	\caption{\label{raman_summary}\label{table_ordering} Summary of the phases of our model, and the point symmetry group to which each one belongs to. We determine whether the point groups break or not the symmetries of D$_{4h}$. The symmetries are  time-reversal  ($\cal{T}$), inversion ($\cal{I}$), rotations of $\pi/2$ (${\cal C}_{4}$) and reflections (${\cal P}_{xy}$) related to $xy$ plane.}
	
	\begin{ruledtabular}
		\begin{tabular}{ccccc}
			Label in Fig.\ref{fig1} & (a) & (b) & (c) & (d) \tabularnewline
			Phase & Paramagnetic  & Local staggered order & Modulated spin liquid & Chiral spin liquid
			\tabularnewline
			Space group & No 139 ($D_{4h}-I4/mmm$) & No 123 ($D_{4h}-P4/mmm$)  & No 126 ($D_{4}-P4/nnc$) & No 128 ($C_{4h}-P4/nmc$) \tabularnewline
			Set of parameters & $\Delta_{0}=\Delta_{m}=\Delta_{c}=0$ & $\Delta_{0}\neq0$, $\Delta_{m}=\Delta_{c}=0$&
			$\Delta_{0}=0$ or $\Delta_{0}\neq0$,&$\Delta_{0}=0$ or $\Delta_{0}\neq0$,\tabularnewline
			&&& $\Delta_{m}\neq 0$ and $\Delta_{c}=0$   &$\Delta_{m}= 0$ and $\Delta_{c}\neq0$
			\tabularnewline
			Lattice & BCT & ST & ST & ST\tabularnewline
			${\cal T}$& yes & yes & yes & no \tabularnewline
			${\cal I}$& yes & yes & no & yes \tabularnewline
			${\cal C}_{4}$ & yes & yes & no & no \tabularnewline
			${\cal P}_{xy}$& yes & yes & no & yes
		\end{tabular}
	\end{ruledtabular}
	
\end{table*}

We start with an effective tight-binding model which can be derived from the $J_1-J_2-J_3$ Heisenberg model on the BCT lattice \cite{Farias2016}. In this model, $J_1$ is the exchange interaction between the nearest-neighbor \textit{interlayer} sites and $J_2$ and $J_3$ are exchange interactions between nearest and next-nearest-neighbor \textit{intralayer} sites, respectively. The ingredient to stabilize the spin liquid phases is the presence of frustration. The phase diagram of this model contains regions of magnetic instability and, in addition, the classical magnetic states have an extreme degeneracy, which prevents magnetic order at $T=0$. By using the slave-boson approach \cite{Pepin2011,Thomas2013}, it is possible to write the spins in terms of complex fermion operators and to define, accordingly, three order parameters with a mean-field decoupling. These fermions thus describe in a unified effective tight-binding model the low energy excitations from the various spin-liquid, magnetic, or paramagnetic ground states \cite{Pepin2011, Thomas2013, Farias2016}.

The different order parameters that define the effective tight-binding model represent three different phases: staggered antiferromagnetic order, modulated spin liquid (MSL) and chiral spin liquid (CSL) \footnote{Ref.\cite{Farias2016} presents the phase diagram from where our effective model is derived. It is shown how SL states can be stabilized in the realistic three-dimensional $J_{1}-J_{2}-J_{3}$ Heisenberg model for the BCT lattice. By using a SU($n$)-symmetric generalization of this Heisenberg model for quantum spin $S$ operators. The phase diagram shows that, for small $n$, the most stable solutions correspond to four different families of long-range magnetic orders that are governed by the couplings $J_1$, $J_2$, and $J_3$. The possible instabilities of these phases are identified for $n=2$, in large $S$ expansions, up to the linear spin-wave corrections. The purely magnetic orders occur for $n\leqslant3$ while SL solutions are stabilized for $n  \geqslant 10$. The SL solution governed by $J_{1}$ breaks the lattice translation symmetry. For $4 \leqslant n \leqslant 9$, it is shown how the competition between $J_1$, $J_2$, and $J_3$ couplings can turn the magnetically ordered ground state into a SL state.}. 
Without loss of generality, we will refer here to the antiferromagnetic order as a commensurate local staggered order (LSO). This effective phenomenological model can also emerge from a realistic microscopic approach when correlations stabilize spin liquid states \cite{Pepin2011, Farias2016} or any other commensurate order with a folded Fermi surface, such as spin interorbital density waves \cite{Riseborough2012}, Fermi surface instability \cite{Oppeneer2011} or spin-density waves \cite{Das2012}. In this work, we do not consider further effects of fluctuations on top of those already mentioned correlated effects. Instead, we focus on the low energy Fermi liquid-like excitations of these states with the specific BCT lattice symmetry breaking.

The effective Hamiltonian, after the mean-field decoupling, assumes the simple form 
\begin{equation}
\centering
H = \sum_{i}(\epsilon_{0}+\Delta_{0}(-1)^{i})c^{\dagger}_{i} c_{i}
+ \sum_{\langle i,j\rangle}t_{ij} c^{\dagger}_{i}c_{j},
\label{hamiltonian1}
\end{equation}
and it describes spinless fermions moving throughout the sites of a BCT lattice by the action of a kinetic hopping term $t_{ij}$. The $c_{i}^{\dagger}$ and $c_{i}$ are the creation  and annihilation operators of spinless fermion at a given site $i$ at position $\bm{R}_{i}$.
Here, these spinless fermions are simply spinons in a deconfined phase \cite{Baskaran1987, Anderson1196,Marston1989,Wen2002}. The parameter $\epsilon_{0}$ adjusts the chemical potential while $\Delta_{0}$ defines the LSO with this local on-site order parameter. Due to the factor $(-1)^{i}$ that multiplies $\Delta_{0}$, the sign alternates between neighboring layers, see  Fig.\ref{AFM}. The hopping $t_{ij}$ connects nearest-neighbours sites on the BCT lattice and breaks the BCT lattice symmetry down to simple tetragonal lattice. It can assume either real or complex values. 

If $t_{ij}$ is complex, we define it as $t_{ij} = t_{1} \pm i \Delta_{c}$ between nearest-neighbors \textit{interlayers}. The imaginary part relates itself to the CSL by the gap parameter $\Delta_{c}$. The sign choice in $t_{ij}$ refers to the income ($+$) or outgoing ($-$) link orientation between two neighboring sites at different planes, see Fig.\ref{csl}. The CSL phase describes loop currents between different \textit{interlayer} sites, as can be seen from Fig.\ref{csl}. If $t_{ij}$ is real, we define it as $t_{ij} = t_{1} \pm\Delta_{m}$  between nearest-neighbors \textit{interlayers}. The gap parameter $\Delta_{m}$ stands for the MSL phase. The choice of plus (dashed red line) or minus (full black line) sign represents the different bonds between neighboring sites at different \textit{interlayers} planes describing, thus, our MSL phase, see Fig.\ref{msl}. Moreover, we assume the presence of homogeneous hopping $t_{2}$ and $t_{3}$ for \textit{intralayer} nearest-neighbors and next-nearest-neighbors sites, see Fig.\ref{hopping}. For the sake of clarity, we do not investigate the possibility of an in-plane order in different lattice symmetry breaking scenarios.

In Table \ref{table_ordering}, we summarize the different spatial groups of each phase of our model. Concerning the space group symmetry analysis of Harima \textit{et al.} \cite{Harima2010}, we start in the framework of a normal paramagnetic state with BCT lattice associated with space group No 139 ($I4/mmm$, D$_{4h}$). The phase transition from PM to HO lowers the lattice symmetry from BCT to ST lattice by making the $\textmd{Z}$ and $\Gamma$ points of the first Brillouin zone of the BCT and ST being connected by a nesting commensurate vector $\bm{Q}$, as signaled by neutron scattering experiments \cite{Bourdarot2010,Villaume2008, Elgazzar2009}. This nesting commensurate vector $\bm{Q}=(1,0,0)$ characterizes the low energy excitations and is the main feature that distinguishes HO from AFM phases of URu$_2$Si$_2$ when pressure is applied. A sharp excitations at $\bm{Q}$ appears in the hidden order phase and disappears when entering the AFM \cite{Bourdarot2010, Villaume2008}. This result is also in agreement with Shubnikov-de Haas measurements \cite{Hassinger2010}. Moreover, the sharp peak obtained by Buhot \textit{et al.} with Raman scattering matches the neutron resonance at $\bm{Q}$ \cite{Buhot2014}. Here, we take $\bm{Q}=(0,0,1)$.

We assume three possible different ST space groups as candidates for the HO. (\textbf{i}) The space group No 123 ($P4/mmm$, D$_{4h}$) characterizes the  LSO phase; (\textbf{ii}) The  space group No 126 ($P4/nnc$, D$_{4}$) for MSL; (\textbf{iii}) the space group No 128 ($P4/mnc$, C$_{4h}$) characterizes the CSL state. Moreover, the selection rules in a Raman spectroscopy experiment for point group D$_{4}$ and C$_{4h}$ have the same selection rules as the initial mother space (point) group D$_{4h}$ \cite{Harima2010}. 

Performing the Fourier transformation of $c_{i}$ operators with $c_{i} = (1/\sqrt{N})\sum_{\bm{k}} \  e^{i \bm{k} \cdot \bm{R}_{i}} c_{\bm{k}},
\label{Fourier_transform}
$ where the sum in $\bm{k}$ runs over the first BZ of the BCT lattice and $N$ is the number of lattice sites, we write down the Hamiltonian in $\bm{k}$-space with a folded BZ from the BCT to the ST (see appendix \ref{appA}), as
\begin{equation}
H= \sum_{\bm{k}}\Psi^{\dagger}_{\bm{k}}{\cal H}({\bm{k}})\Psi_{\bm{k}}.\label{hamiltonian2}
\end{equation}
However,  the sum in $\bm{k}$ now runs over the first BZ of the ST lattice. By using the definition of $\Psi_{\bm{k}}= (c_{\bm{k}}, c_{\bm{k}+\bm{Q}})^{T}$, we define ${\cal H}(\bm{k})$ as
\begin{equation}
{\cal H}({\bm{k}})=\left(\begin{array}{cc}
\varepsilon({\bm{k}}) & \Delta({\bm{k}})\\
\Delta^{\dagger}({\bm{k}}) & \varepsilon({\bm{k}+\bm{Q}})
\end{array}\right),
\label{E_matrix}
\end{equation}
where $\epsilon(\bm{k})=\epsilon_{0}+t_{1} \gamma_{1}(\bm{k})+t_2 \gamma_{2}(\bm{k})+t_3 \gamma_{3}(\bm{k})$ is the dispersion in one band and $\Delta(\bm{k})= \Delta_{0} \pm i \Delta_{SL}f_{SL}(\bm{k})$ is the gap between different bands connected by the wave vector $\bm{Q}$. The parameter $\Delta_{SL}$ refers either to a modulated ($\Delta_m$) or to a chiral ($\Delta_c$) spin liquid and the plus or minus sign is for CSL or MSL, respectively. We define the structure factor $f_{SL}(\bm{k})$ for each SL phase as 
\begin{eqnarray}
f_{c}(\bm{k})&=&\, 8\sin\left(\frac{k_{x}a}{2}\right)\sin\left(\frac{k_{y}a}{2}\right)\cos\left(\frac{k_{z}c}{2}\right),\label{dispersionCSL}\\
f_{m}(\bm{k}) &=&\, 8\sin\left(\frac{k_{x}a}{2}\right)\sin\left(\frac{k_{y}a}{2}\right)\sin\left(\frac{k_{z}c}{2}\right)\label{dispersionMSL},
\end{eqnarray}
The other structure factors are
\begin{eqnarray}
\gamma_{1}(\bm{k})&=&\; 8 \cos\left(\frac{k_{x}a}{2}\right)\cos\left(\frac{k_{y}a}{2}\right)\cos\left(\frac{k_{z}c}{2}\right),\\
\gamma_{2}(\bm{k})&=&\; 2 (\cos\left(k_{x}a\right)+\cos\left(k_{y}a\right)),\\
\gamma_{3}(\bm{k}) &=&\; 4 \cos\left(k_{x}a\right)\cos\left(k_{y}a\right),
\label{structure_factors}
\end{eqnarray}
with $a$ and $c$'s being the ST lattice constants (for derivations of the above equation see appendix \ref{appA}).
The diagonalization of the effective Hamiltonian in Eq.(\ref{E_matrix}) gives the dispersion relations
\begin{equation}
\resizebox{0.5\textwidth}{!}{$
	E^{\pm}(\bm{k})= \frac{\epsilon(\bm{k})+\epsilon(\bm{k}+\bm{Q})}{2} \pm \sqrt{ |\Delta(\bm{k})|^{2} + \left(\frac{\epsilon(\bm{k})-\epsilon(\bm{k}+\bm{Q})}{2}\right)^{2}}$
}
\end{equation} 
which are defined in the ST Brillouin zone.  

The existing symmetries of our model are time-reversal ${\cal T}$, inversion ${\cal I}$, four-fold ${\cal C}_4$ ($\pi/2$) rotations and two reflection symmetries with respect to the planes ${\cal P}_{x/y}$. The ${\cal T}$ invariance requires ${\cal{T}}^{-1} H(\textbf{k}){\cal T}=H(-\textbf{k})$, which means that $\Delta({\bm{k}})=\Delta^{\dagger}({-\bm{k}})$. This condition is satisfied in the MSL case. For the CSL phase, we automatically break the time-reversal symmetry. Concerning the ${\cal C}_4$ ($\pi/2$) rotations, both MSL and CSL break rotational symmetry, in accordance with torque measurements \cite{Okazaki2011}. The MSL phase breaks the reflection symmetry with respect to the plane ${\cal P}_{xy}$ while CSL preserves that rotation symmetry.


\section{The $A_{2g}$ symmetry 	\label{secIII}}

In this section, we turn to the experimental signatures of the phases presented in the previous section when considering a Raman scattering experiment. We are particularly interested in the response at $A_{2g}$ symmetry. 

In a Raman experiment, the $A_{2g}$ symmetry is accessed by polarization directions that transform like $xy (x^{2} -y^{2})$ and it is usually associated with chiral excitations \cite{Devereaux2007}. This symmetry has been studied in the context of high-temperature superconductors \cite{Shastry1990, Khveshchenko1994, Salamon1995} where the denomination of magnetic Raman scattering refers to the scattering by the spin degrees of freedom and to fluctuations of a chiral operator \cite{Shastry1990, Khveshchenko1994}. This type of scattering is dominated by resonant contributions which are produced by the current operator instead of the effective density operator.

The distinction between the non-resonant and resonant regimes is associated with the energy scales of the light used in the experiment. The former case occurs when the energy of the incident or scattered photon is smaller if compared to the energy gap between different bands in the system, otherwise we refer to the latter case \cite{Shastry1990,Devereaux1997}. Moreover if there is no exchange in the momentum, i.e., if the difference between the initial and final momentum goes to zero as $\bm{q}\rightarrow0$, the effective-mass approximation \cite{Ashcroft2011} is well suited to calculate the Raman vertex \cite{Devereaux1997}. However, the effective mass approximation is unable to produce a signal for the $A_{2g}$ unless we have unusual electronic band structure with $\partial^{2}E(\bm{k})/\partial k_x \partial k_y \neq \partial^{2}E(\bm{k})/\partial k_y \partial k_x $ in some region of $\bm{k}$-space on the Fermi surface \cite{Buhot2014}. In the next subsection, we present our calculation for the current-current correlation function.

\subsection{Current-current correlation function}

In order to obtain signatures of $A_{2g}$ symmetry, we consider a different vertex from the one given by the effective mass approach. We use the idea put forward by Khveshchenko and Wiegmann in the context of high-energy large shift Raman scattering in Mott insulators systems \cite{Khveshchenko1994}. These authors assume that the energies of incident and scattered light are much larger than the widths of relevant electronic bands,and thus, we are dealing with a resonant process. In this limit, the Raman cross section can be written in terms of the ground state correlation function of the scattering tensor $\hat{M}^{A_{2g}}$. This scattering tensor is proportional to the equal-time current-current commutator.  

We define the scattering operator in terms of a current-current commutator at equal time that in the framework of our non-interacting effective model essentially is given by
\begin{equation}
\hat{M}_{\mu\nu}(\bm{q},\tau) = [\hat{j}_{\mu}(\bm{q},\tau),\hat{j}_{\nu}(\bm{q},\tau)].
\label{scattering_op}
\end{equation}
where $\hat{M}$	 is the scattering tensor, $\mu$ ($\nu$) $= x,y,z $, $\bm{q}$ is the momentum exchanged during the scattering process, and $\tau$ is the imaginary time. The current operator $\hat{j}$ is written as 
\begin{equation}
\hat{j}_{\mu}(\bm{q},\tau)=\sum_{\bm{k}}
\Psi^{\dagger}_{\bm{k}+\bm{q}/2}(\tau)
\tilde{\bm{\gamma}}^{\mu}_{R}(\bm{k})
\Psi_{\bm{k}-\bm{q}/2}(\tau).
\label{current_operator2}
\end{equation} 
Here, the $\tilde{\bm{\gamma}}^{\mu}_{R}(\bm{k})$ is the matrix that originates from the derivatives of ${\cal H}(\bm{k})$ with respect to a momentum component $k_{\mu}$. The index $R$ refers to the resonant contribution.  
We can extract the signal for $A_{2g}$ symmetry by computing the correlation function
\begin{equation}
	\tilde{\chi}^{A_{2g}}\left(\bm{q},i\omega_{m}\right)=-\frac{1}{\cal V}\int_{0}^{\beta}d\tau\:e^{i\omega_{m}\tau}\langle T_{\tau}\hat{M}^{A_{2g}}(\bm{q},\tau)\hat{M}^{A_{2g}}(\bm{-q},0)\rangle.
\end{equation}	
The $T_{\tau}$ is the imaginary time-ordering operator, ${\cal V}$ is the volume of the BZ, and the scattering operator $\hat{M}^{A_{2g}}(\bm{q},\tau)= \hat{M}_{xy}(\bm{q},\tau)-\hat{M}_{yx}(\bm{q},\tau)$.  For other symmetries, like $A_{1g}$ ($x^{2}+y^{2}$), $B_{1g}$ ($x^{2}-y^{2}$) and $B_{2g}$ ($xy$), the scattering operator defined in Eq.(\ref{scattering_op}), involves other combinations of $x$ and $y$ directions. At this point we can make $\bm{q}\rightarrow0$ without loss of generality, and, then, we find that
\begin{widetext}
\begin{equation}
\tilde{\chi}^{A_{2g}}(i\omega_{m}) = \frac{1}{\beta {\cal V}} \sum_{\bm{k},n}\\  	\textnormal{Tr} [ \tilde{\gamma}^{A_{2g}}_{R}(\bm{k})  G(\bm{k};i\nu_{n}+i\omega_{m})     
\tilde{\gamma}^{A_{2g}}_{R}(\bm{k})  
G(\bm{k};i\nu_{n})].
\label{A2g_correlation}
\end{equation}
\end{widetext}

We use the finite temperature representation of Matsubara Green's functions \cite{Mahan}, where $\beta=1/k_{B}T$ is the usual inverse of the temperature, and $\omega_{m}=2m\pi/\beta$ is the bosonic Matsubara frequency. We perform the analytical continuation in the frequency domain  $i\omega_{m} \rightarrow \omega + i\delta$, and extract the imaginary part of the correlation function as the Raman response. The parameter  $\delta$ is defined as a small scattering rate.
The Green functions are expressed in terms of the elements of  eigenbasis, i.e.  $G^{\pm}(\bm{k};i\nu_{n})= 1/(i\nu_{n}- E^{\pm}(\bm{k}))$. The $E(\bm{k})$ is the eigenvalues introduced in Sec.\ref{secII} and $\nu_{n}=(2n+1)\pi/\beta$ are the fermionic Matsubara frequencies. As a results, we obtain a new vertex for the $A_{2g}$ symmetry that is different from the effective mass approximation \cite{Devereaux1997}
\begin{equation}
\tilde{\bm{\gamma}}^{A_{2g}}_{R}(\bm{k})=\frac{\partial {\cal H}(\bm{k})}{\partial k_{x}}\frac{{\cal H}(\bm{k})}{ \partial k_{y}} - \frac{\partial{\cal H}(\bm{k})}{\partial k_{y}}\frac{\partial{\cal H}(\bm{k})}{\partial k_{x}}.\label{a2g2}
\end{equation}

The above definition is similar to the commutation relation between the first derivatives of $ {\cal H}(\bm{k})$ with respect to the $k_{x}$ and $k_{y}$ components. Notice that, $\tilde{\bm{\gamma}}^{A_{2g}}_{R}(\bm{k})$ can be nonzero, because $H({\bm k})$ is given by Eq.(\ref{E_matrix}) and is a $2\times 2$ matrix (see Appx.\ref{appB}). Since the resonant Raman process requires a fourth order vertex \cite{Devereaux1997}, by inspecting our correlation function in Eq.(\ref{A2g_correlation}), it is easy to verify that this requirement is automatically satisfied. In contrast, by using the definition of the scattering operator $\hat{M}_{\mu\nu}$ in Eq.(10) the final vertex representation for other irreducible symmetries like $A_{1g}$, $B_{1g}$ and $B_{2g}$ vanish identically. In principle, the other symmetries responses could also be calculated using the effective mass approximation [8]. However, we did not address this question here, since our goal is to compare our results with the Raman experiment and to arrive at the $A_{2g}$ response.

\subsection{Numerical calculations}

In this section, we evaluate numerically the correlation function given by Eq.(\ref{A2g_correlation}). Our numerical results were obtained by using multidimensional Monte Carlo $\bm{k}$-integration over the the first Brillouin zone of the simple tetragonal lattice with the open routine of GNU Scientific Library \cite{Galassi2009}. Here, we take the zero temperature limit, i.e. $T\rightarrow0$, and keep $t_{1}$ non-zero and equals to $1.0$ to set the scale for other parameters. This value of $t_{1}$ leads to a bandwidth of $8t_{1}$ meV, which guarantees that we are using the correct energy scale compatible with the HO temperature $T_{0}\approx17.5$ K. The factor of eight results from the BCT lattice coordination number. This particular value of the bandwidth is for the spinless fermions in our effective model. In a more realistic calculation with real heavy conduction electrons, these bandwidth can be even larger. However, to account for such an effect would require a Kondo-like Hamiltonian, as described Ref.\cite{Montiel2014}. This analysis is outside the scope of this present work. Furthermore, for the sake of simplicity, we assume $\epsilon_{0}=0$, since this parameter do not show any quantitative or qualitative influence in our results.
\begin{figure}
	\centering	\includegraphics[height=7cm,width=9.0cm]{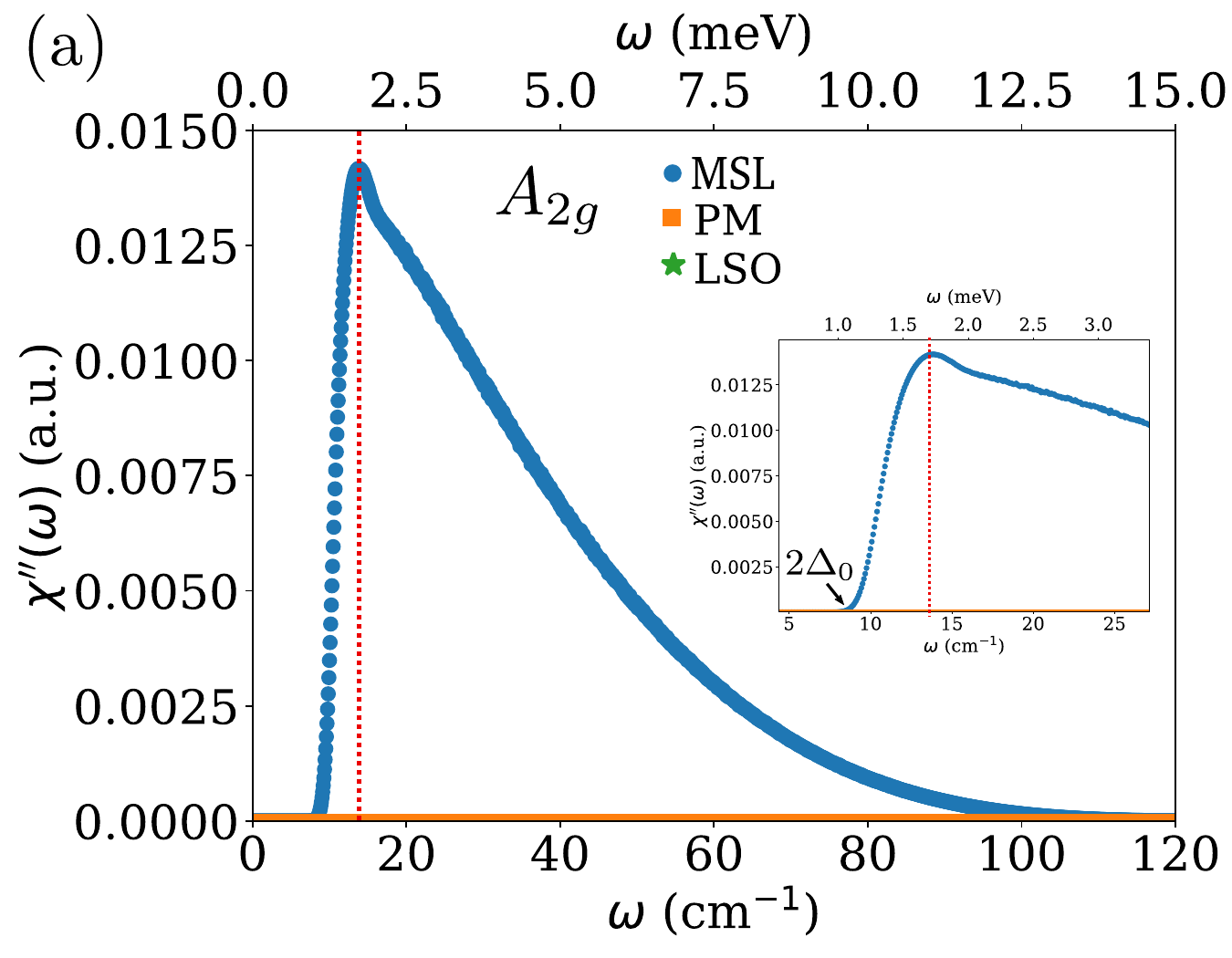}\\
	\bigskip
	\includegraphics[height=7cm,width=9.0cm]{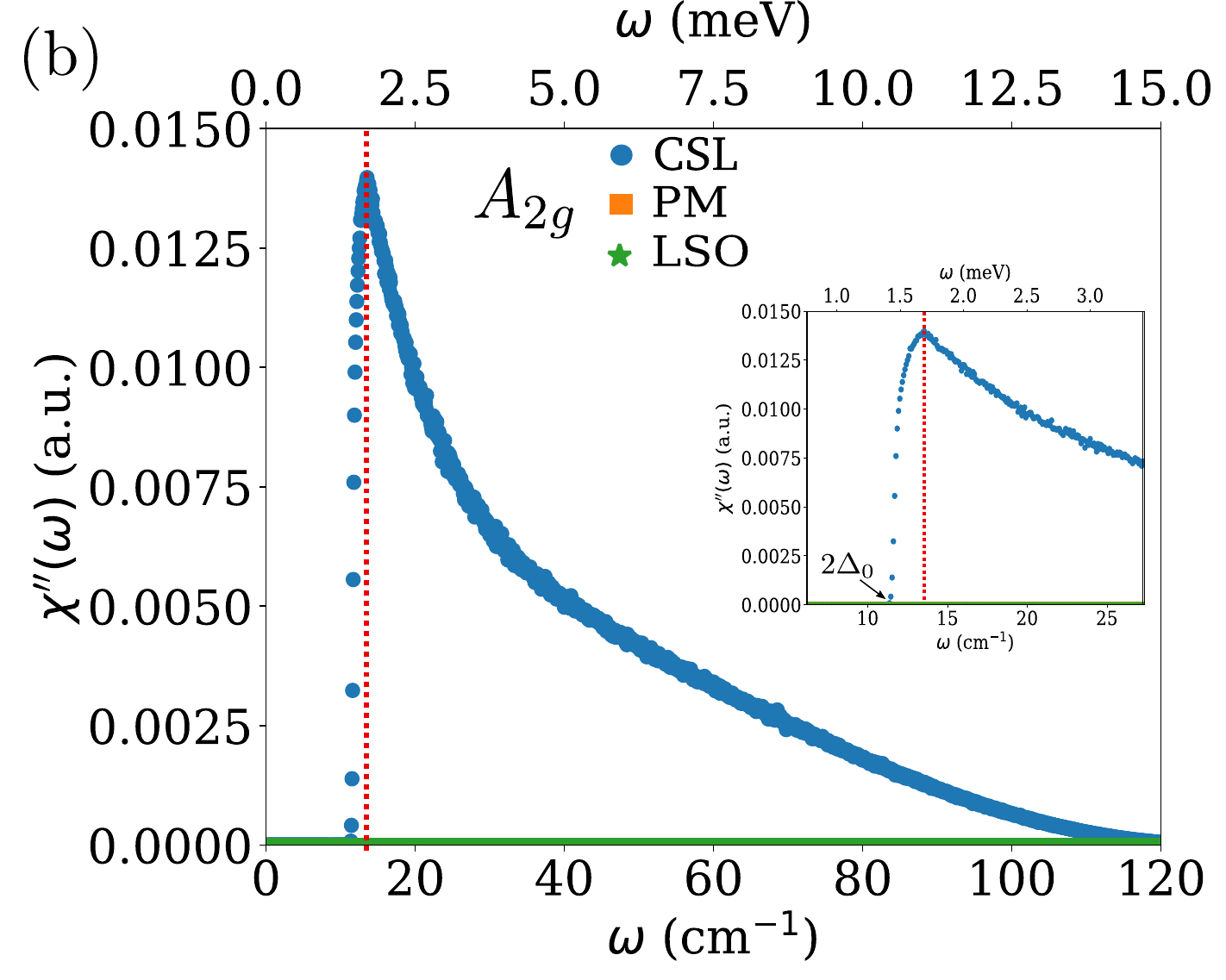}
	\caption{\small{The Raman response in $A_{2g}$ symmetry displays an inelastic peak for (a) modulated spin liquid and (b) chiral spin liquid phases. We have set the parameter in (a) $\Delta_{m}=0.1$ and $\Delta_{0}=0.6$ while in (b) $\Delta_{c}=0.065$ and $\Delta_{0}=0.72$, such that the maximum of the peak would coincide approximately at $\omega\approx1.7$ meV (insets) that is the experimental value of frequency at which the peak was observed. The response is observed only for nonzero values $\Delta_{SL}$, as we can see for the PM (square orange points) and LSO (star green points) where $\chi''$ is identically zero. Non zero values of $\chi''$ start to appear for $\omega=2\Delta_{0}$, for the MSL and CSL phase.}}
	\label{fig4}
\end{figure}
 
Fig.\ref{fig4}-(a) and \ref{fig4}-(b) show the Raman response for the modulated spin liquid and the chiral spin liquid, respectively.
In both cases, the Raman response in the $A_{2g}$ channel produces an inelastic peak which appears precisely at $\omega=2\Delta_{0}$, for nonzero values of parameter $\Delta_{0}$ and one of $\Delta_{SL}$. When we take only the hopping $t_{1}$ (PM phase), we do not produce any signal as well as when we take finite values of $t_{1}$ and $\Delta_{0}$ (LSO phase), square orange and asterisks green points, respectively. These two curves are superimposed on the zero value of $\chi''(\omega)$. The maximum value of the peak increases as we increase the spin liquid parameters. 

\begin{figure}[!t]
	\includegraphics[height=8cm,width=9cm]{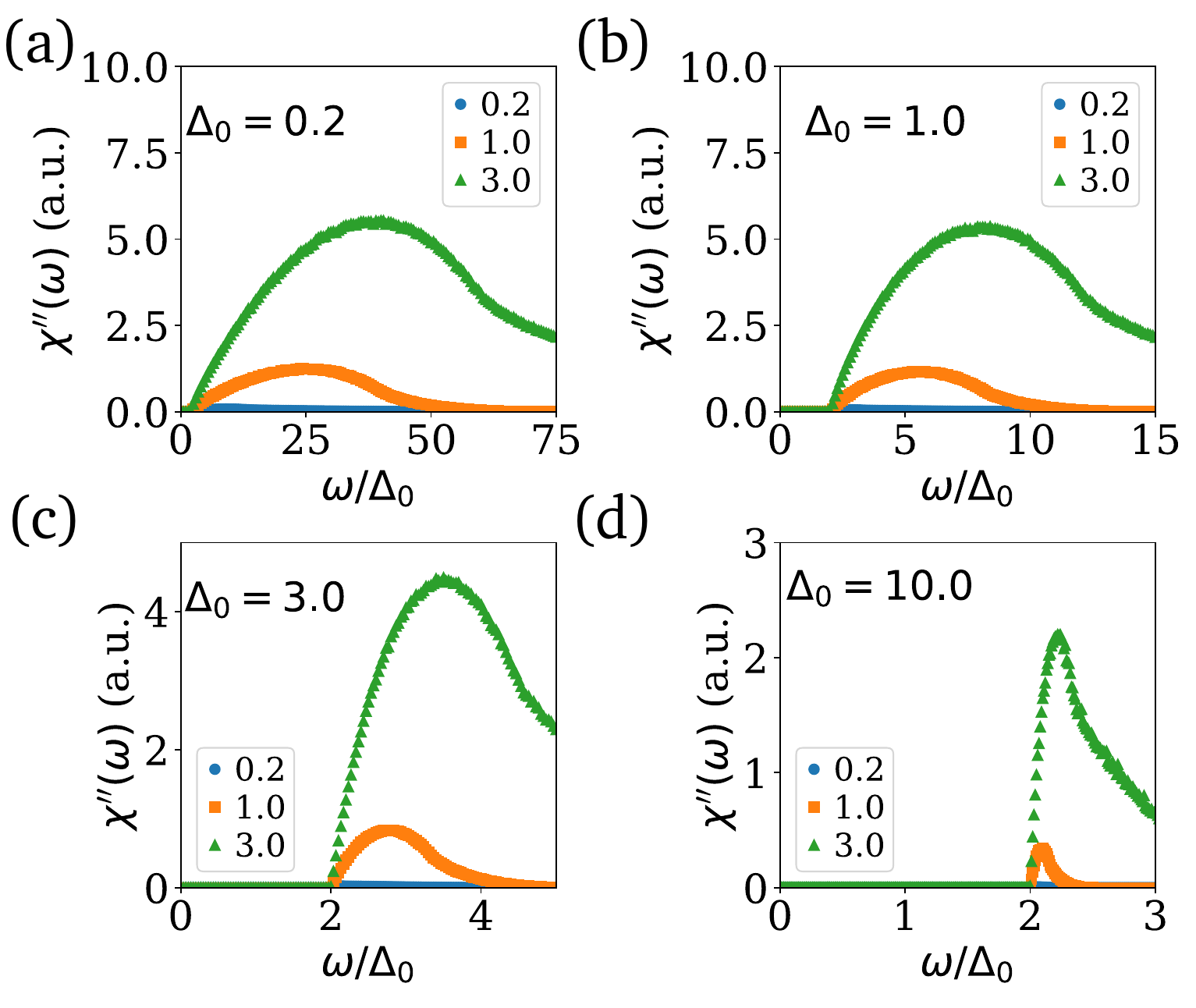}
	\caption{\small{Raman response $A_{2g}$ symmetry in the MSL phase that corresponds to space group No 126. $\Delta_{m}$ is varied from $0.0$ up to $3$, while kept $\Delta_{0}$ fixed to \textbf{(a)} 0.2, \textbf{(b)} 1.0, \textbf{(c)} 3.0  and \textbf{(d)} 10. We found a inelastic peak that starts to emerge at $\omega=2\Delta_{0}$, as can be clearly seen from plots \textbf{(c)} and \textbf{(d)}.}}
	\label{fig2}
\end{figure}

\begin{figure}[!t]
	\includegraphics[height=8cm,width=9cm]{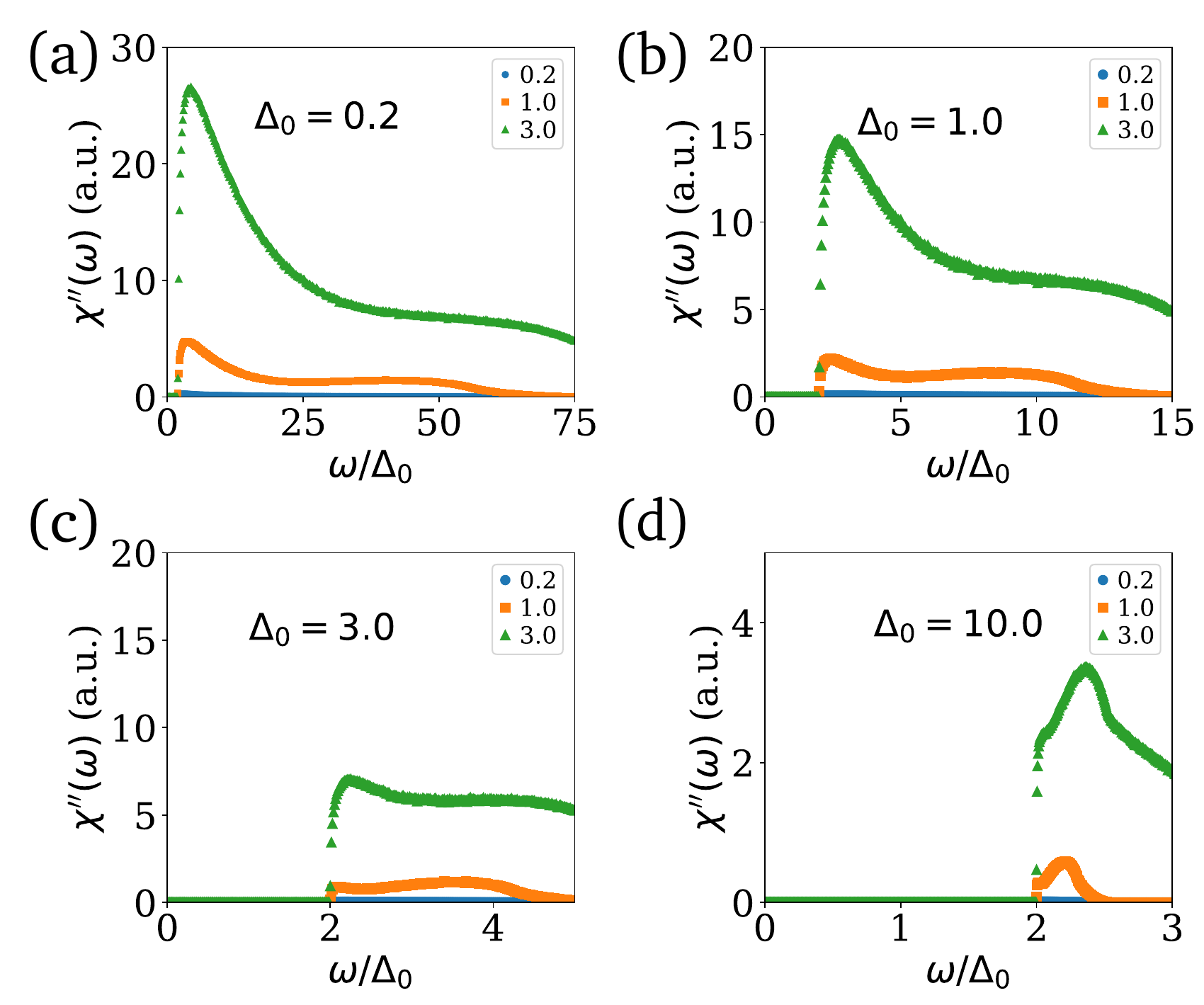}
	\caption{\small{Raman response in $A_{2g}$ symmetry for CSL, which corresponds to space group No 128. As above, we varied the values of $\Delta_{c}$ from $0.0$ up to $3$, while kept $\Delta_{0}$ fixed to \textbf{(a)} 0.2, \textbf{(b)} 1.0, \textbf{(c)} 3.0  and \textbf{(d)} 10. Again, the inelastic peak  starts to appear at $\omega=2\Delta_{0}$.}}
	\label{fig3}
\end{figure}

In Fig.\ref{fig2} and Fig.\ref{fig3}, we show that the inelastic peak starts to manifest itself precisely at $\omega=2\Delta_{0}$. In both cases, we demonstrate that there is a relationship between the parameters $\Delta_{0}$ and $\Delta_{SL}$ in producing the Raman response. 
In sub-figures (a)-(d), we vary the values of $\Delta_{SL}$, while keeping $\Delta_{0}$ constant. For $\Delta_{SL}=0.2$, and whatever $\Delta_0$, the Raman response is small, of the order of $10^{-2}$, as shown in Fig.\ref{fig4}; this is a much smaller response than what is obtained for $\Delta_{SL}=1$ or $3$, as presented in Fig.\ref{fig2} and \ref{fig3}. These calculations allow us to extract the specific roles of each parameter. In principle, $\Delta_{SL}$ is responsible for the maximum value of the peak. In addition, $\Delta_{0}$ controls the peak width being responsible for making the peak broader or narrower, i.e., the smaller its value, the broader is the peak. These qualitative conclusions were useful for the correct adjustment of parameters in Fig.\ref{fig4}(a)-(b) in order to obtain a qualitative agreement with the data from URu$_2$Si$_2$. This naturally lead us  us to choose $\Delta_{0}=0.6$, for the  MSL and $\Delta_{0}=0.72$ for the CSL with the maximum of the peak being manifested at $\omega \approx 1.7$ meV, while choosing to keep a small value for $\Delta_{SL}$ in both cases.


\subsection{Comparison with UR\lowercase{u}$_{2}$S\lowercase{i}$_{2}$ \label{subsecC}}

The electronic Raman scattering data \cite{Buhot2014,Kung2015} showed that there is an electronic continuum at $6.8$ meV for the $A_{2g}$ symmetry. Below this value of frequency, a sharp excitation peak appears inside the electronic gap.  At the phase transition to the hidden order phase, the gap opens up. In comparison with these experimental results, we find a non-zero signal in $A_{2g}$ channel, both in the MSL and CSL phases but not in purely LSO phase.
On the other hand, our calculations does not produce any signal in other symmetries than $A_{2g}$. The Raman response in other channels will be produced by the usual effective mass approximation \cite{Devereaux2007}. This analysis based on the effective mass approach is somehow meaningful. However, we did not address this issue here, and  focus on the $A_{2g}$ response only.

The Raman signature we obtain for $A_{2g}$ symmetry is in qualitative agreement with the inelastic signal that was experimentally  observed by Buhot \textit{et. al} \cite{Buhot2015}. In Fig.\ref{fig4}(a)-(b), we show this inelastic peak for specific values of $\Delta_{0}$, $0.6$  in the MSL phase, and $0.72$ for CSL phase. With this value, we can adjust the peak to appear around at $1.7$ meV, which is the experimental value where the sharp excitation was observed. Our peak $A_{2g}$ is qualitatively consistent with the experimental results, although there is an intrinsic broadening from our calculations making our peak wider than that observed experimentally. However, we have to keep in mind that our theoretical calculation is obtained from an effective tight-binding model. The consideration of supplementary interaction terms within a standard Random Phase Approximation from our non-interacting result could naturally provide a narrowing of this peak.

The origin of a inelastic peak suggests the presence of well-defined quasiparticles. In the framework of the cuprates systems \cite{Liu1993}, the $A_{2g}$ symmetry is associated with the appearance of elementary excitations in which quantum fluctuations destroy the N\'eel order. These elementary excitations have a strongly chiral nature \cite{Khveshchenko1994, Wen1989, Wen2002}. Such a scenario favors the choice of chiral spin liquid as the natural candidate for the appearance of such excitations. Moreover, in their analysis, Harima \textit{et al.} \cite{Harima2010} suggested that the space group No 128 is incompatible with the nuclear resonance experiments on URu$_2$Si$_2$.  Additionally, space group No 128 is compatible with the commensurate chirality density wave state discussed in Kung \textit{et al.} \cite{Kung2015}. In contrast, the proposition of a modulated spin liquid \cite{Pepin2011,Thomas2013} as a candidate for HO is based on breaking of a four-fold rational symmetry which was detected in the HO phase from susceptibility measurements \cite{Okazaki2011}. Furthermore, a modulated spin liquid scenario is also compatible with other experimental investigations of HO phase, such as, Hall data \cite{Oh2007}, Thermal conductivity \cite{Behnia2005} and quantum oscillations \cite{Nakashima_2003}. From our study, we cannot attest which phase is the best to describe HO, since our results show no distinction between the MSL and CSL.


Since the signal is already observed when one of the parameters for spin liquid are active together with the $\Delta_{0}$, this fact clearly eliminates the purely commensurate LSO state as the main source for the Raman signal. Moreover, it suggests that there might be at least an interplay between the LSO phase and a MSL or CSL phase. In fact, propositions of a super-vector order parameter, which accounts for the interplay between HO and the large moment antiferromagnetic (LMAF) phase in the URu$_2$Si$_2$ have been already put forward \cite{Chandra2013, Haule2009}. However, without considering such sophisticated phenomenological  models, we provide in a relatively simple effective theory a plausible description that the HO and LMAF phases are intricately connected with each other, although, our results take into account an specific experimental signature provided by Raman scattering experiments. In other words, here our results favors this interplay scenario.

On the other hand, the electronic continuum at $A_{2g}$ symmetry is not reproduced. The inclusion of effects of fluctuation of the parameters involved that characterize each phase in our model might remedy this problem. In addition, to make it clear that the phases considered here are indeed spin liquid states one could compute spin-spin correlations to provide that both spin liquid present some decay in the spin correlations which indicates the absence of long-range order. Another experimental possibility is to investigate what happens to the $A_{2g}$ under pressure. In this respect, a full track of the $A_{2g}$ signal could be obtained from the HO to the AFM phase. We expect that this investigation of the Raman signal under applied pressure reproduce a signal owing to the commensurate local staggered order only, which means that in this phase a suppression of the peak in the $A_{2g}$ would be observed. However, we leave these complementary analysis for future work.

Finally, we note that our work is relevant in the context of recent Raman experiments which had not been considered in the point group symmetry analysis by Harima \textit{et al} \cite{Harima2010}. For instance, a recent work in which the authors reported nuclear magnetic resonance measurements on URu$_2$Si$_2$ \cite{Harima2018} and is based on a group theory analysis has concluded that the hidden order state belongs to the point group $P4/nnc$ and the electronic state is a rank five odd-parity electronic dotriacontapolar order with a commensurate ordering wave vector $\bm{Q}=(0,0,1)$. This result indicates that a phase similar to the modulated spin liquid \cite{Pepin2011,Thomas2013}  might be the strongest candidate to describe this mysterious phase of matter.

 
\section{Conclusions \label{secIV}}

This paper presented an effective model to describe Raman scattering experiments in systems that display the BCT lattice. In the process, we investigated the connection of two different spin liquid states in a BCT lattice structure the emergence of a Raman signal at $A_{2g}$ symmetry which was experimentally observed in URu$_2$Si$_2$ \cite{Buhot2014, Kung2015}. 

In particular, the essential ingredient for generating a Raman signal in the $A_{2g}$ symmetry is the interplay between a local staggered order,  which can be in principle associated with the AFM observed in URu$_{2}$Si$_{2}$ under pressure, and a modulated or chiral spin liquids phases, that could describe the HO state. In all cases, we reproduce the $A_{2g}$ feature \cite{Buhot2014, Kung2015} observed with Raman scattering experiments, and found that both spin liquids (modulated and chiral) present similar Raman responses by displaying an inelastic peak. Lastly, a clear distinction between the narrow peak and the gap is still to be addressed in a future work of $A_{2g}$ symmetry.

\begin{acknowledgments}
We acknowledge discussion with I. Paul, S. Magalhães, J. Buhot, R. G. Pereira, V. S. de Carvalho, H. Harima, C. Pépin, C. Lacroix, D. Baeriswyl, and E. Miranda. C. Silva de Farias thanks the finantial support from international cooperation program CAPES/COFECUB and CNPq. This work was partially supported by the ANR-DFG grant Fermi-NESt.
\end{acknowledgments}

%




\appendix

\section{The dispersion relations \label{appA}}
The purpose of this appendix is to clarify the derivations the two structure factors $f_{m}(\bm{k})$ and $f_{c}(\bm{k})$ for the MSL and CSL, respectively.
We take the Hamiltonian in Eq.(\ref{hamiltonian1})
\begin{equation}
H = \sum_{i}(\epsilon_{o}+\Delta_{0}(-1)^{i})c^{\dagger}_{i} c_{i}
+ \sum_{\langle i,j\rangle}t^{}_{ij} c^{\dagger}_{i}c_{j}.
\label{hamiltonian}
\end{equation} 
The sum in lattice sites is associated with the BCT lattice. The Hamiltonian has two part $H=H_{0}+H_{t}$, where $H_{0}=\sum_{i}(\epsilon_{o}+\Delta_{0}(-1)^{i})c^{\dagger}_{i} c_{i}$ and $H_{t}=\sum_{\langle i,j\rangle}t^{}_{ij} c^{\dagger}_{i}c_{j}$.
Let us concentrate on the hopping term
\begin{equation}
H_{t}=\sum_{\langle i,j\rangle}t^{}_{ij} c^{\dagger}_{i}c_{j}.
\end{equation}
We carried out the calculation for a general hopping $t_{i,j}$ and at the end we will only concentrate on the first inter-plane neighbors, since the in-plane relations can be derived in the same fashion as in square lattice. For interplane hopping, we have that
\begin{eqnarray}
t_{ij} &=&  t_{1} \pm i \Delta_{c},
\end{eqnarray}	
and
\begin{eqnarray}
t_{ij} &=&  t_{1} \pm \Delta_{M}.
\end{eqnarray}
The sign plus or minus take into account the nature of the link between two sites, as explained in section \ref{secII}. With the Fourier transform in Eq.(\ref{Fourier_transform}), we rewrite the Hamiltonian and we end up with
\begin{eqnarray}
H=\frac{1}{N}\sum_{\bm{k,k'}}c_{\bm{k}}^{\dagger}c_{\bm{k'}}\sum_{i}e^{-i\bm{(k-k')\cdot R_{i}}}E_{\bm{R}_{i}}(\bm{k}'),\nonumber\\
\end{eqnarray}
where $E_{\bm{R}_{i}}(\bm{k}')=\sum_{\delta}t_{i,i+\delta}e^{i\bm{k}'\cdot \bm{\delta}}$ with $j=i+\delta$, wherw $\delta$ is a first neighbor vector. The sum in $\bm{k}$ runs over the first Brillouin zone of BCT lattice. The next step is to split the sum in the BCT-BZ  as being a sum of two tetragonal lattices, where we assumed that the BCT is bipartite in two sub-lattices A and B. We end up with four terms 
\begin{eqnarray}
H & = & 	\frac{1}{N}\sum_{\bm{k,k'}}c_{\bm{k}}^{\dagger}c_{\bm{k}'}\sum_{i}e^{-i(\bm{k}-\bm{k}')\cdot \bm{R}_{i}}E_{\bm{R}_{i}}(\bm{k}') \nonumber\\
& + &\frac{1}{N}\sum_{\bm{k,k'}}c_{\bm{k}}^{\dagger}c_{\bm{k'}}\sum_{i}e^{-i\bm{(k-k')\cdot R_{i}}}E_{\bm{R}_{i}}(\bm{k}'+\bm{Q})\nonumber\\
&+&\frac{1}{N}\sum_{\bm{k,k'}}c_{\bm{k}+\bm{Q}}^{\dagger}c_{\bm{k}'}\sum_{i}e^{-i\bm{(k+Q-k')\cdot R}_{i}}E_{\bm{R}_{i}}(\bm{k}')\nonumber\\
&+&\frac{1}{N}\sum_{\bm{k,k'}}c_{\bm{k}}^{\dagger}c_{\bm{k+Q}}\sum_{i}e^{-i\bm{(k-Q-k')\cdot R}_{i}}E_{\bm{R}_{i}}(\bm{k}'+\bm{Q}).\nonumber\\
\end{eqnarray}
Now the sums in $\bm{k}$ run over the T-lattice, while the sums in $i$ run over the BCT lattice.

We can also split the sum in the BCT-Lattice in a sum for sub-lattice A and  B.	This requires that $t_{i,i+\delta}=t_{\delta}^{A} $ if $\bm{R}_{i}\in$ A or $t_{i,i+\delta}=t_{\delta}^{B} $ if $\bm{R}_{i}\in$ B.	The condition $t_{\delta}^{B}=\left(t_{\delta}^{A}\right)^{\ast}$ preserves the hermiticity of the Hamiltonian, which give us the relation between $E_{B}(\bm{k}')= (E_{A}(\bm{k}'))^{\ast}$. We know  that $\frac{1}{N}\sum_{\bm{R}_{i}}e^{-i\bm{(k-k')}\cdot \bm{R}_{i}}=\frac{1}{2}\delta_{\bm{k}\bm{k}^{'}}$ and $e^{-i\bm{Q}\cdot \bm{R}_{i}}= \pm 1$ if $\bm{R}_{i}$ belongs to A or B, respectively. Finally, we write that
%
\begin{eqnarray}
H \	&=& \ \frac{1}{2}\sum_{\bm{k}}c_{\bm{k}}^{\dagger}c_{\bm{k}}[E(\bm{k})+E^{\ast}(\bm{k})]\nonumber\\
&+&	\frac{1}{2}\sum_{\bm{k}}c_{\bm{k+Q}}^{\dagger}c_{\bm{k}+\bm{Q}}[E(\bm{k}+\bm{Q})+E^{\ast}(\bm{k}+\bm{Q})]\nonumber\\
&+&	\frac{1}{2}\sum_{\bm{k}}c_{\bm{k}+\bm{Q}}^{\dagger}c_{\bm{k}}[E(\bm{k})-E^{\ast}(\bm{k})]\nonumber\\
&+&	\frac{1}{2}\sum_{\bm{k}}c_{\bm{k}}^{\dagger}c_{\bm{k+Q}}[E(\bm{k+Q})-E^{\ast}(\bm{k}+\bm{Q})].
\end{eqnarray}
%
\begin{figure}
\includegraphics[height=5.4cm]{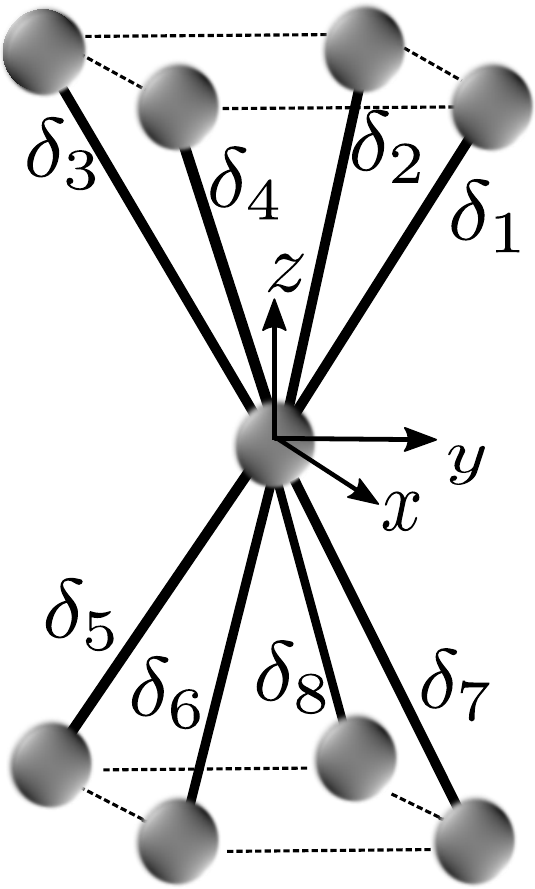}
\caption{The BCT lattice with its eight first neighbors interplane. The values of each unitary vector are $\boldsymbol{\delta}_{1}=(a/2,a/2,c/2)$,$\boldsymbol{\delta}_{2}=(-a/2,a/2,c/2)$, $\boldsymbol{\delta}_{3}=(-a/2,-a/2,c/2)$,
	$\boldsymbol{\delta}_{4}=(a/2,-a/2,c/2)$,
	$\boldsymbol{\delta}_{5}=-\boldsymbol{\delta}_{1}$, $\boldsymbol{\delta}_{6}=-\boldsymbol{\delta}_{2}$, $\boldsymbol{\delta}_{7}=-\boldsymbol{\delta}_{3}$, and $\boldsymbol{\delta}_{8}=-\boldsymbol{\delta}_{4}$.	
\label{neighbours}}
\end{figure}

In matrix representation, it follows that 
\begin{eqnarray}
H &=& \sum_{\bm{k}}\Psi^{\dagger}_{\bm{k}}h'_{\bm{k}}\Psi_{\bm{k}},
\end{eqnarray}
with the definition of $\Psi_{\bm{k}}= (c_{\bm{k}}, c_{\bm{k}+\bm{Q}})^{t}$ and
\begin{eqnarray}
h'_{\bm{k}}&=&\left(\begin{array}{cc}
\frac{E(\bm{k})+E^{\ast}(\bm{k})}{2} &\frac{E\left(\bm{k+Q}\right)-E^{\ast}\left(\bm{k}+\bm{Q}\right)}{2} \\
\frac{E(\bm{k})-E^{\ast}(\bm{k})}{2}	 &\frac{E\left(\bm{k+Q}\right)+E^{\ast}\left(\bm{k}+\bm{Q}\right)}{2}
\end{array}\right).
\label{E_matrix1}
\end{eqnarray}

By taking the definition of $E(\bm{k})$, the hopping $t_{i,j}$ and considering the wave vector $\bm{Q}=\{1,1,1\}$, we can recognize the following terms when doing the sum in $\delta$ neighbors vectors
\begin{equation}
\frac{E(\bm{k})+E^{\ast}(\bm{k})}{2} =  t_{1} \gamma_k^{1}+t_2 \gamma_k^{2}+t_3 \gamma_k^{3},
\end{equation}
\begin{equation}
\frac{E\left(\bm{k+Q}\right)+E^{\ast}\left(\bm{k}+\bm{Q}\right)}{2}	= - t_{1} \gamma_k^{1}+t_2 \gamma_k^{2}+t_3 \gamma_k^{3},
\end{equation}
\begin{equation}
\frac{E\left(\bm{k+Q}\right)-E^{\ast}\left(\bm{k}+\bm{Q}\right)}{2} = \pm i 8\Delta_{SL}f_{SL}(\bm{k}),
\end{equation}
\begin{equation}
\frac{E(\bm{k})-E^{\ast}(\bm{k})}{2} = \mp i 8 \Delta_{SL}f_{SL}(\bm{k}).
\end{equation}
Note the change in the choice of sign on the last two equations as a consequence of the ordering wave vector $\bm{Q}$. If we add the contribution from $\Delta_{0}$ and $\epsilon_{0}$ we recover the ${\cal H}(\bm{k})$ matrix in Eq.(\ref{E_matrix}) defined in section \ref{secII}.

We would like to highlight the derivation of $\bm{k}$ dependence in $\Delta(\bm{k})$. We take the first neighbors interlayer and, in this case, the sum in $\delta$ takes in to account eight different neighbors interlayer in the BCT lattice represented in Fig.\ref{neighbours}. The first sum in $\delta$ for $t_{1}$ hopping produces the $\gamma^{1}_{k}$ factor and the second sum produces the $f_{SL}(\bm{k})$ function, which can assume $f_{c}(\bm{k})$ or $f_{m}(\bm{k})$, as defined in Eq.(\ref{dispersionCSL}) and (\ref{dispersionMSL}), depending on the definition of $t_{ij}$. This complete our demonstration of the relation presented in section \ref{secII}.
\section{The $A_{2g}$ vertex \label{appB}}

In this appendix, we show how to extract a non-zero response in the $A_{2g}$ symmetry. The expression of $A_{2g}$ vertex was defined in section \ref{secIII} and is written as
\begin{equation}
\tilde{\bm{\gamma}}^{A_{2g}}(\bm{k})=\frac{\partial {\cal H}(\bm{k})}{\partial k_{x}}\frac{\partial {\cal H}(\bm{k})}{ \partial k_{y}} - \frac{\partial {\cal H}(\bm{k})}{\partial k_{y}}\frac{\partial {\cal H}(\bm{k})}{\partial k_{x}}.
\label{a2g3}
\end{equation}
This expression is analogous to the following commutator
\begin{equation}
\tilde{\bm{\gamma}}^{A_{2g}}(\bm{k})=\left[\frac{\partial {\cal H}(\bm{k})}{\partial k_{x}},\frac{\partial {\cal H}(\bm{k})}{\partial k_{y}}\right].
\label{commutation_a2g}
\end{equation}
We must emphasize that ${\cal H}(\bm{k})$ is a matrix and not a pure function of momentum $\bm{k}$. Therefore, the commutator defined above does not vanish. To go further, we can make use of the Pauli matrices $\sigma^{\alpha}$ ($\alpha=1,2,3$) together with the identity, and rewrite ${\cal H}(\bm{k})$ as 
\begin{equation}
{\cal H}(\bm{k})=A({\bm{k}})\mathbbm{1}
+\Re[\Delta(\bm{k})]\sigma^{1}+\Im[\Delta(\bm{k})]\sigma^{2}+ B({\bm{k}})\sigma^{3},
\end{equation}
where $\mathbbm{1}$ is the identity matrix, and
\begin{eqnarray}
A({\bm{k}})=\frac{\epsilon({\bm{k}})+\epsilon({\bm{k}+\bm{Q}})}{2},\\
B({\bm{k}})=\frac{\epsilon({\bm{k}})-\epsilon({\bm{k}+\bm{Q}})}{2}.
\label{parametros}
\end{eqnarray}
From this point and on, we change the notation for the imaginary part $\Im[\Delta(\bm{k})]\equiv\Delta''_{\bm{k}}$, as well as the real part $\Re[\Delta(\bm{k})]\equiv\Delta'_{\bm{k}}$, and use the commutation relations of Pauli matrices to rewrite the commutator as
%
\begin{align}
\left[\frac{\partial {\cal H}(\bm{k})}{\partial k_{x}},\frac{\partial {\cal H}(\bm{k})}{\partial k_{y}}\right] &=
2i\left(\frac{\partial \Delta''_{\bm{k}}}{\partial k_{x}}\frac{\partial B({\bm{k}})}{\partial k_{y}}-\frac{\partial B({\bm{k}})}{\partial k_{x}}\frac{\partial \Delta''_{\bm{k}}}{\partial k_{y}}\right)
\sigma^{1}\nonumber\\
&+
2i\left(\frac{\partial B({\bm{k}})}{\partial k_{x}}\frac{\partial \Delta'_{\bm{k}}}{\partial k_{y}}-\frac{\partial \Delta'_{\bm{k}}}{\partial k_{x}}\frac{\partial B({\bm{k}})}{\partial k_{y}}\right)\sigma^{2}\nonumber\\
&+
2i\left(\frac{\partial \Delta'_{\bm{k}}}{\partial k_{x}}\frac{\partial \Delta''_{\bm{k}}}{\partial k_{y}} -	\frac{\partial \Delta''_{\bm{k}}}{\partial k_{x}}\frac{\partial \Delta'_{\bm{k}}}{\partial k_{y}}\right)\sigma^{3}.
\end{align}
%
In our model, we explicitly define the form of $\Delta({\bm{k}})$. With this definition, all the contributions that involve derivatives of the real part vanishes identically, because it is a constant defined by the parameter $\Delta_{0}$. The only term that survives is the first one on the right-hand side. Therefore, we have 
\begin{equation}
\tilde{\bm{\gamma}}^{A_{2g}}(\bm{k})= 2i\left(\frac{\partial \Delta''_{\bm{k}}}{\partial k_{x}}\frac{\partial B_{\bm{k}}}{\partial k_{y}}-\frac{\partial B_{\bm{k}}}{\partial k_{x}}\frac{\partial \Delta''_{\bm{k}}}{\partial k_{y}}\right)
\sigma^{1}
\end{equation}

From this result, we can check if the $\gamma^{A_{2g}}(\bm{k})$ is invariant under time-reversal symmetry. For spinless particles, the time-reversal operator ${\cal T}$ is directly connected to the complex conjugation $K$, i.e., ${\cal T}=K$. If a particular operator is invariant under time-reversal symmetry, this means that
\begin{eqnarray}
{\cal T}\hat{A}{\cal T}^{-1}= \hat{A}. 
\end{eqnarray}
For our vertex $\gamma^{A_{2g}}({\bm{k}})$, it follows that 
%
\begin{eqnarray}
{\cal T}\tilde{\bm{\gamma}}^{A_{2g}}(\bm{k}){\cal T}^{-1}= -	\gamma^{A_{2g}}({\bm{k}})
\end{eqnarray}

However, time-reversal operation also changes momentum from $\bm{k} \rightarrow -\bm{k}$. Here we keep in mind the definitions of $B({\bm{k}})$ in Eq.(\ref{parametros}) and $\Delta({\bm{k}})= \Delta_{0} \pm i \Delta_{SL}f_{SL}(\bm{k})$. The function $B({\bm{k}})$ is always even because the  dispersion $\epsilon({\bm{k}})$ is even. For the imaginary part of $\Delta({\bm{k}})$, we consider the dispersions from the two spin liquids. For the chiral spin liquid, $f_{{c}}({\bm{k}})$ is even while for the modulate spin liquid, $f_{{m}}({\bm{k}})$ is odd. Therefore, we conclude that
\begin{eqnarray}
f_{c}(-\bm{k})=	f_{c}(\bm{k}),\\
f_{m}(-\bm{k})=	-f_{m}(\bm{k}).
\end{eqnarray}
These results show that in terms of time-reversal symmetry only the modulated spin liquid has a $\tilde{\bm{\gamma}}^{A_{2g}}(\bm{k})$ which is truly time-reversal invariant. In contrast, the chiral spin liquid phase breaks time-reversal symmetry as expected from the definitions of what chiral phases should be \cite{Wen1989}.

\bibliographystyle{apsrev4-1_control}
\bibliography{./Ref}

\end{document}